\documentclass[%
 reprint,
 amsmath,amssymb,
 aps,
]{revtex4-2}

\usepackage{subfigure,amsfonts}
\usepackage{siunitx}
\usepackage{graphicx}
\usepackage{dcolumn}
\usepackage{bm}
\usepackage[dvipsnames]{xcolor}

\begin{document}


\title{Momentum transfer in the ponderomotive potential of near-infrared laser pulses leads to sizable energy shifts and electron-wavepacket squeezing in  time-resolved ARPES}

\author{Xinwei Zheng}%
\affiliation{Fachbereich Physik, Freie Universität Berlin, Arnimallee 14, 14195 Berlin}%

\author{Martin Weinelt}%
\affiliation{Fachbereich Physik, Freie Universität Berlin, Arnimallee 14, 14195 Berlin}%

\author{Christian Strüber}
\email{christian.strueber@fu-berlin.de}
 \affiliation{Fachbereich Physik, Freie Universität Berlin, Arnimallee 14, 14195 Berlin}

\date{\today}

\begin{abstract}
We observe momentum transfer in the ponderomotive potential of near-infrared (NIR) laser pulses in time- and angle-resolved photoemission spectroscopy (tr-ARPES) experiments with ultrashort extreme ultraviolet probe pulses. Acceleration of photoelectrons in the transient grating provided by an intense laser pulse reflected at a surface leads to delay-dependent oscillations of electric kinetic energies. Photon and electron momenta determine the oscillation frequency. We experimentally observe and theoretically simulate electron yield modulations driven by a novel electron-energy bunching effect. Measurement results are simulated and fitted with high accuracy. Complete reversion of the ponderomotive momentum transfer allows for retrieval of the undisturbed initial state and the transient band structure for overlapping pump and probe pulses.



\end{abstract}

\maketitle


\section{\label{Introduction} Introduction }

Angle-resolved photoelectron spectroscopy (ARPES) is one of the most important experimental methods to study the electronic properties of solids. It provides insight to the band structure and Fermi surface of solids by using detectors with energy and angular resolution \cite{hufner_photoelectron_2003}. In time- and angle-resolved photoelectron spectroscopy (tr-ARPES) the dynamical response of the electronic structure is studied. A first pump pulse excites the system, which is photo-ionized at a certain time delay by a subsequent probe pulse. Mapping the transient response of electronic states and electron and spin populations the temporal evolution of the system and the underlying microscopic interactions is investigated \cite{MWeinelt_2025}. Typically, the photoemission spectra at negative delays, i.e. the probe pulse arrives first, are assumed to be comparable to the unperturbed ground state. However, with intense pump pulses space charge effects, in which Coulomb interaction with electrons that are emitted through high-order above-threshold photoemission, lead to an energy redistribution \cite{ZHOU200527}. Furthermore, interaction of emitted electrons with the light field of the pump pulse leads to a modification of the detected spectra. Famously, it is possible to use intense ultrashort carrier-envelope-phase stable laser pulses in attosecond streaking experiments to accelerate photoelectrons in the vector potential of a laser pulse and gain precise temporal information \cite{krausz_attosecond_2009,cavalieri_attosecond_2007,Okell:15,siek_angular_2017}. Recently the subcycle emergence of Floquet-Bloch bands from surface states on a topological insulator was investigated with mid-infrared driving fields in tr-ARPES \cite{Ito2023}.\par

\begin{figure}[tbhp]
	\centering
	\includegraphics[width=0.8\linewidth]{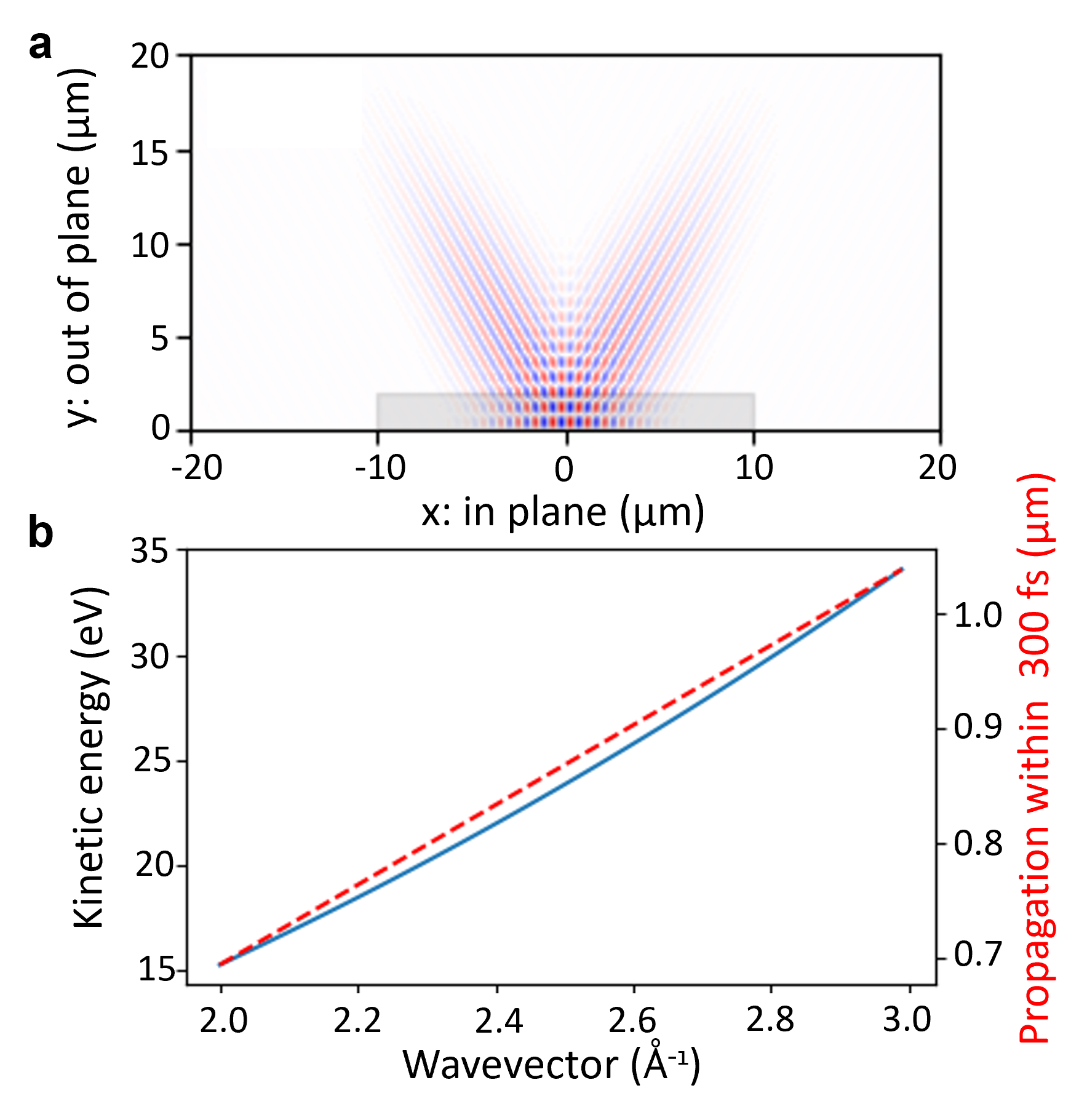}
	\caption{(\textbf{a}) Electric field of an ultrashort laser pulse that is reflected at a surface with an incidence angle of 60°. Marked gray area indicates the region shown in Fig~\ref{Mom_standing_wave}. (\textbf{b}) Photoelectron kinetic energy and propagation distance within 300 fs in dependence of wave vector.}
	\label{Fig1}
\end{figure}

\begin{figure}[tbhp]
	\centering
	\includegraphics[width=0.95\linewidth]{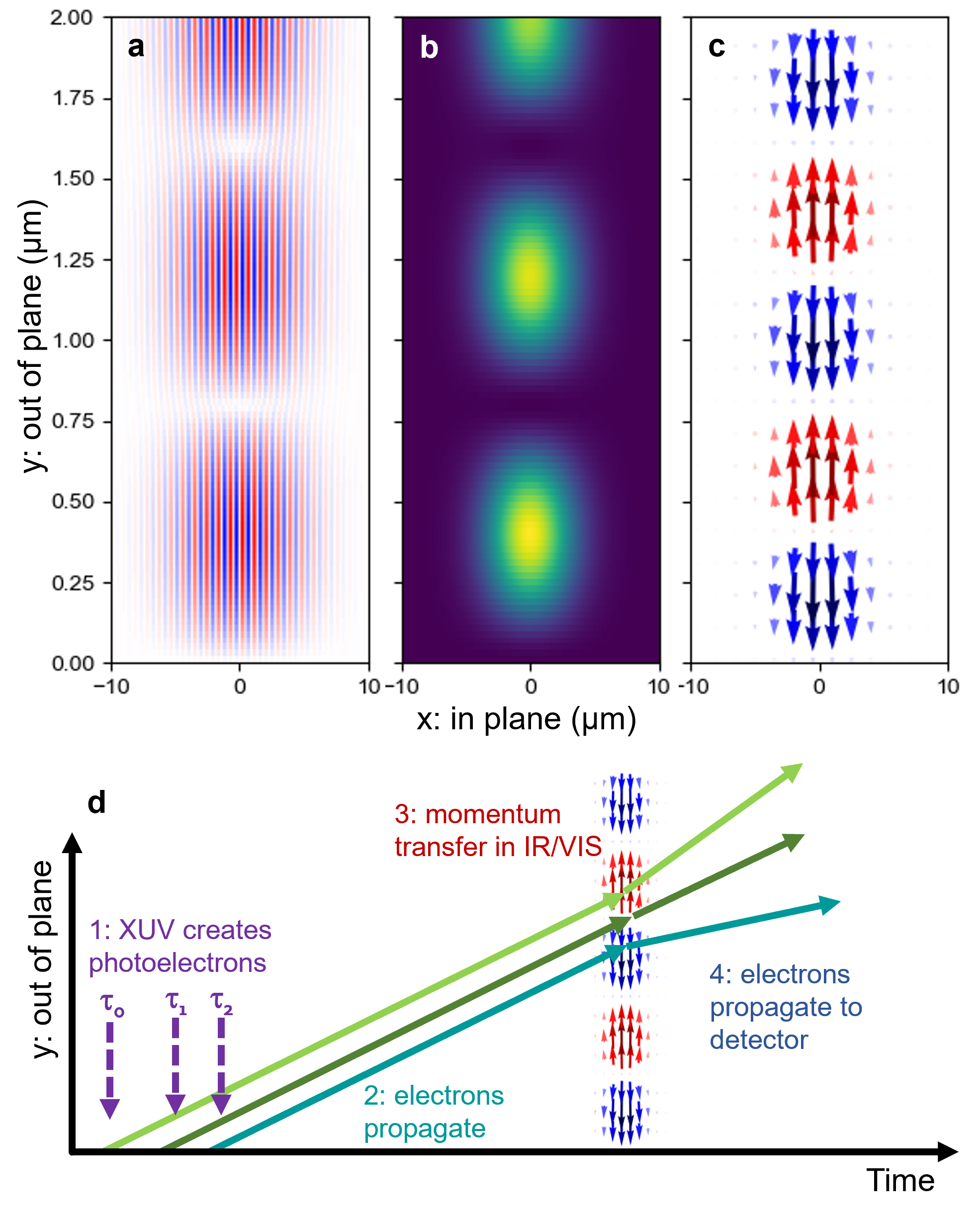}
	\caption{ Electric field (\textbf{a}) and intensity (\textbf{b}) of an ultrashort laser pulse that is reflected at a surface. (\textbf{c}) Force on an electron due to intensity gradient. (\textbf{d}) Sketch of main steps in ponderomotive momentum transfer (1) XUV pulse creates photoelectron. Three arrows indicate electrons started at three different XUV delay positions ($\tau_0$,$\tau_1$,$\tau_2$)  (2) Photoelectrons travel with identical velocity away from the surface before the arrival of an ultrashort laser pulse. (3) Momentum transfer due to light intensity gradient creating a force on charged particles. (4) Electrons propagate towards detector with individually altered velocity. }
	\label{Mom_standing_wave}
\end{figure}

In most ARPES setups the pump pulses have much lower field amplitudes and, even more importantly, are polarized in the sample plane, preventing acceleration in the laser field. A light-intensity driven effect was demonstrated by Bovensiepen et al. in 2009 \cite{bovensiepen_ultrafast_2009}. They reported an oscillatory modulation of the kinetic energy of photoelectrons from the Gd(0001) surface when the infrared pump pulse arrives later than the ultraviolet probe pulse. Nevertheless, a kinetic energy variation with a small amplitude of 0.8\,meV was observed in the photoemission spectrum generated by the ultraviolet probe pulses with a photon energy of 6\,eV. It was explained by a momentum transfer process involving the ponderomotive acceleration of the emitted photoelectrons in a transient grating formed by the interference between the incident and reflected parts of the pump pulse. By analyzing the momentum change of photoelectrons, it is possible to address ultrafast charge-carrier dynamics of metals \cite{bovensiepen_ultrafast_2009}.
In Figure~\ref{Fig1}\textbf{a}, the electric field of a laser pulse reflected on a perfect mirror surface is shown for an incidence angle of 60°. In and around the gray shaded area, the still incident and the already reflected parts of the pulse interfere and form a standing wave pattern. The photoelectrons' wavevector and kinetic energy determine the propagation away from the surface (see Fig.~\ref{Fig1}\textbf{b}). Figure~\ref{Mom_standing_wave}  shows the basic principles of the interaction. The electric field exhibits nodes in parallel and perpendicular to the surface (see Fig.~\ref{Mom_standing_wave}\textbf{a}). Please note that the axis aspect ratio are changed. The plot limits are corresponding to the gray shaded area in  Fig.~\ref{Fig1}\textbf{a}. The intensity is mainly modulated perpendicular to the surface (see Fig. \ref{Mom_standing_wave}\textbf{b}). Consequently, the resulting force on the electron is opposite to the intensity gradient and points towards or away from the surface (see Fig.~\ref{Mom_standing_wave}\textbf{c}). When XUV pulses arrive first and generate photoelectrons, those travel away from the sample until the arrival of the intense infrared or visible laser pulse. The electron position in the transient standing wave determines whether the electron is accelerated or decelerated due to the light intensity gradient creating a force on it. As the standing wave exhibits a node on or close to the surface, depending on the reflection phase shift of the material, electrons experience no considerable momentum transfer in the temporal overlap of XUV and laser pulse. However, with increasing optical delay and electron propagation distance the momentum transfer exhibits an oscillatory behavior.\par 

In recent years, the development of bright coherent sources of extreme-ultraviolet (XUV) photons enabled to investigate electrons with higher momentum \cite{MWeinelt_2025,frietsch_high-order_2013}. Time-resolved ARPES with XUV photon energies is capable to map transient band structures in the whole Brillouin zone of solids.  Even though the ponderomotive acceleration model predicts larger kinetic energy variations at higher probe photon energies \cite{bovensiepen_ultrafast_2009}, there have been no reports of momentum transfer in the ponderomotive potential of the pump pulse except for our study, where the effect was briefly mentioned \cite{K_Bobowski_Sci_Adv_2024}.\par

Time-resolved ARPES is an important and powerful technique to study the signature of ultrafast quasiparticle dynamics and collective excitations in the electronic structure (for a recent review see \cite{MWeinelt_2025}). It is, \textit{e.g.}, used to monitor demagnetization processes in magnetic materials ranging from itinerant $3d$ magnets \cite{Rhie2003,tengdin_critical_2018} to rare earth materials where $4f$ electrons are carrying most of the magnetic moments \cite{Bovensiepen2007,carley_femtosecond_2012,Frietsch2020,K_Bobowski_Sci_Adv_2024} as well as to investigate charge-density-wave and exciton dynamics in 2D materials \cite{Merboldt2025,Reutzel31122024,PhysRevLett.125.096401,doi:10.1126/sciadv.adk3897,Choi2025,PhysRevB.107.115136} or collective excitations in superconductors \cite{Zhang2014}. The combination with spin detection methods enables to investigate the spin properties and dynamics of dispersed electronic states with high energy and time resolution \cite{andres_separating_2015,eich_band_2017,gort_early_2018}. Efficient optical demagnetization often requires a high pump fluence resulting in adverse space charge effects in the ARPES experiment. Using longer pump wavelengths allows us to maintain high pump fluences by increasing the nonlinear order of the  above-threshold photoemission process. Increasing the pump fluence and the pump wavelength are both expected to scale up the energy modulation as a result of momentum transfer in the ponderomotive potential.\par 

Here, we systematically investigate the influence of momentum transfer in the ponderomotive potential of near-infrared (NIR) laser pulses on photoelectrons emitted from the Fe(110) surface by ultrashort XUV pulses recorded with a hemispherical photoelectron analyzer. In comparison to Ref.~\onlinecite{bovensiepen_ultrafast_2009} we observe up to 30 times larger amplitudes of energy  oscillations. Furthermore, given our angle-resolved method, we are tracing the effect of momentum transfer trough a substantial part of the Brillouin zone of Fe. We clearly observe a dependence of the oscillation frequency on the perpendicular momentum of the electrons. Thus, additional to the kinetic energy, the electron emission angle determines the oscillatory behavior. We observe a reshaping of spectral features that increases with separation of pump and probe pulses and ultimately leads to an energy bunching in the detected spectra. We explain this effect by the increased dephasing of the oscillatory momentum transfer in adjacent parts of the continuous ARPES spectra.\par 
We support our experimental observations by simulations that are able to reproduce even minor features in the tr-ARPES data. Including the full ARPES information into the calculations allows for identifying and explaining the previously not described effect of electron momentum bunching. Fits are performed with high fidelity over a three dimensional data set. Obtained parameters of the momentum transfer can be used by the same numerical routine to reverse the effect and render the undisturbed initial state.\par

\section{\label{Introduction_simulation} Calculations of Momentum Transfer in tr-ARPES}

\subsection{\label{Introd_Analytic} Theoretical Background and Scaling Laws for Momentum Transfer}

In a classical picture charged particles in an electric field oscillating with frequency $\omega$ perform a quiver motion. This motion increases the cycle-averaged particle energy by the ponderomotive energy (or potential)
\begin{equation}\label{eq_U_p}
    U_{P} =  \frac{e^2E^2}{4m\omega^2}= \frac{I}{\omega^2} \cdot \frac{2e^2}{c \epsilon_0 m} 
\end{equation}
Here, we follow the analytical modeling in Ref. \cite{bovensiepen_ultrafast_2009}. As discussed there the gradient of the ponderomotive potential exerts a force $\mathbf{F}_p$ on the charged particle and accelerates it. The total momentum transfer is given by the integral 
\begin{equation}\label{eq_delta_k_integral}
    \hbar \Delta \mathbf{k} =  \int \mathbf{F}_p dt = - \int \nabla U_p dt
\end{equation}
If the intensity of the field is spatially structured or the fields are confined the force can become substantial. When laser pulses are (partially) reflected from a surface an intensity standing wave pattern similar to a transient grating of pulses with crossing paths is created. The light intensity modulation wavelength $\Lambda$ is determined by $k^{ph}_\perp$, the photon wave vector perpendicular to the surface:

\begin{equation}\label{Lambda_standing_Wave}
\Lambda=\frac{2\pi}{2 k^{ph}_\perp}=\frac{\lambda}{2 \cos{\alpha}}=\frac{\pi c}{ \omega \cos{\alpha}}
\end{equation}

The propagation velocity of the standing wave pattern along the surface is proportional to the wavevector component along the surface, and therefore, from the laser wavelength $\lambda$ and the angle of incidence $\alpha$. This implies short intensity oscillations and a stationary pattern in normal incidence and long oscillations and a quickly traveling pattern in gracing incidence.\par

The observed oscillation frequency $\Omega$ is, in the non-relativistic regime, given by the perpendicular electron velocity $v_\perp=\frac{\hbar k_\perp}{m_e}$  and $\Lambda$:

\begin{equation}\label{Oscillation_freq}
\Omega=2\pi \frac{v_\perp}{\Lambda}=2\pi \frac{\hbar k_\perp}{m_e} \cdot \frac{\omega \cos{\alpha} }{\pi c}
\end{equation}
%

The perpendicular momentum of the photoelectron $k_\perp $ is dependent on the kinetic energy $E_{kin}$ and parallel momentum $k_\parallel$ or alternatively the emission angle with respect to the sample surface normal $\beta$ by:

\begin{equation}\label{k_Pythagoras}
k_\perp =  \sqrt{\frac{2m_eE_{kin}}{\hbar^2}-k_\parallel^2}= \sqrt{\frac{2m_eE_{kin}}{\hbar^2}} \cdot \mathrm{cos}(\beta)
\end{equation}
Hence, in tr-ARPES momentum transfer varies both with $E_{kin}$ and $k_\parallel$. Especially at higher parallel momentum a detuning of the oscillation frequency between electrons with identical kinetic energies becomes apparent.\par

With a few considerate assumptions and approximations it is possible to find analytical solutions that describe the strength of the momentum transfer process for gaussian shaped pulses with temporal standard deviation $\sigma$ \cite{bovensiepen_ultrafast_2009}:

\begin{equation}\label{BovenKineticOscillation}
\Delta E_{kin}(\Delta t)=\epsilon e^{-A-B\Delta t^2}\sin(\Omega\Delta t+\psi) \, ,
\end{equation}

where the amplitude is 

\begin{equation}\label{eq:amplitude_CS}
\epsilon=\frac{\Omega}{\omega}\cdot\frac{\Phi\sqrt{R}}{\omega}\cdot\frac{e^2}{\varepsilon_0 m_e c}
\end{equation} 
The incident fluence of the IR pulse is $\Phi=\frac{1}{2}\epsilon_0cE_0^2\cdot\sqrt{2\pi}\sigma$. The full width at half maximum is $2\sqrt{2\ln 2}\sigma$. The reflectivity $R$ and the phase shift $\psi$ of the IR light at the metal surface can be calculated using Fresnel equations with the angle of incidence $\alpha$. For electrons at a given electron momentum $k_\perp $ the ratio $\Omega/\omega$ between oscillation frequency and light frequency only depends on the incidence angle $\alpha$ (see Eq.~\ref{Oscillation_freq}). Hence, Eq.~\ref{eq:amplitude_CS} can be rearranged to

\begin{equation}\label{eq:amplitude_XZ}
\epsilon=\Phi \lambda k_\perp  \cos{\alpha} \sqrt{R} \frac{\hbar e^2}{\pi \varepsilon_0 m_e^2 c^3}\,.
\end{equation} 

\par The amplitude $\epsilon$ is initially damped by
\begin{equation}
A=\frac{2\sigma^2\omega^2\hbar^2 k_\perp^2\cos^2{\alpha}}{(m_ec)^2}=\frac{\sigma^2\Omega^2}{2} \, ,
\end{equation} 
 where $\sigma\Omega$ states the fraction of the oscillation period that is sampled by a finite pulse. Furthermore, there is a decrease with delay by 
\begin{equation}B=\frac{\hbar^2 k_\perp^2\cos^2{\alpha}}{2(m_ec\sigma)^2}=\frac{v_\perp^2}{c^2}\frac{\cos^2{\alpha}}{2\sigma^2}= \frac{1}{2} \left( \frac{\Omega/\omega}{2\sigma} \right)^2 \end{equation} 

From Eqs.~(\ref{BovenKineticOscillation}) and (\ref{eq:amplitude_XZ}) it is possible to derive a simple scaling law for ultrashort pulses where $e^{-A}\simeq1$:
\begin{equation}\label{Scaling_rel}
\Delta E_{kin} \simeq \Phi \lambda  \sqrt{E_{kin}} \cdot \mathrm{cos}(\alpha)  
\end{equation}

Thus, the momentum transfer increases with the initial electron kinetic energy as well as the fluence and wavelength of the laser pulse. Consequently, time-resolved ARPES utilizing ultrashort XUV probe and intense NIR pump pulses is pushing all three parameters towards higher oscillation amplitudes. \par\smallskip

The equations above are giving information on the behavior of a single electron with well defined timing and momentum. However, a tr-ARPES experiment involves finite pump and probe pulses with electron distributions which are quasi-continuous in energy and momentum. For this reason in the following section calculations are expanded to treat a full ARPES spectrum as input and determine the whole tr-ARPES signal of the ponderomotive oscillations.\par

\subsection{\label{Introd_Numeric} Simulation of momentum transfer induced electron bunching in tr-ARPES}

\begin{figure}[tbhp]
	\centering
	\includegraphics[width=0.85\linewidth]{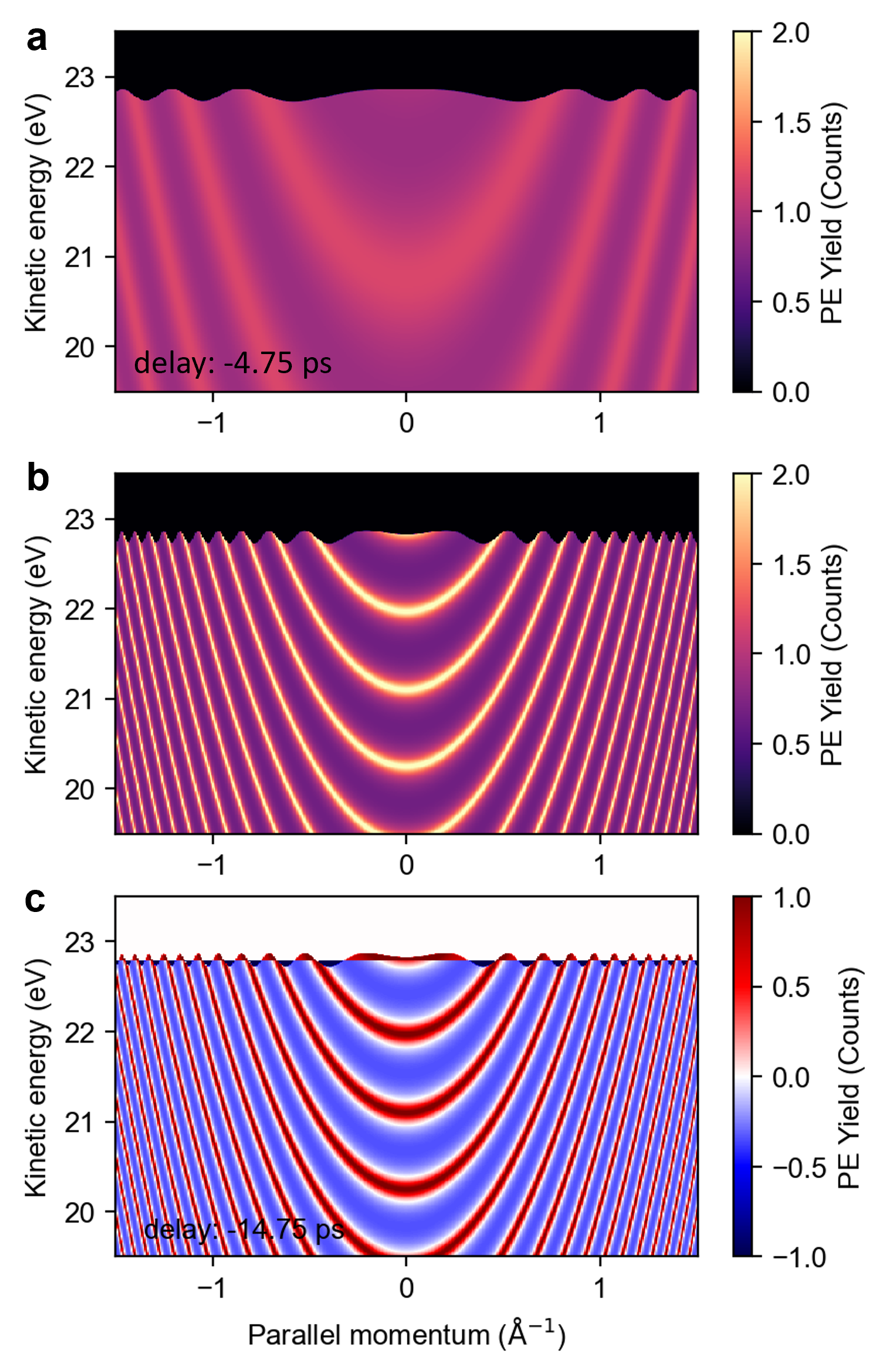}
	\caption{Simulation of ponderomotive momentum transfer starting with a uniform density of states and $E_{\mathrm{F}}=22.8\,\mathrm{eV}$. Simulated ARPES yield after delays $\tau= -4.25\,\mathrm{ps}$ (\textbf{a}) and $\tau= -14.75\,\mathrm{ps}$ (\textbf{b}). (\textbf{c}) Differences to photoemission yield without momentum transfer for $\tau= -14.75\,\mathrm{ps}$.}
	\label{Sim_en_bunching}
\end{figure}

In a simulation of ponderomotive momentum transfer based on the analytic solution given by Eq.~\ref{BovenKineticOscillation} the dynamical changes to an ARPES spectrum with uniform density of state up to an arbitrarily set Fermi energy of $E_{\mathrm{F}}=22.8\,\mathrm{eV}$ are investigated. Input parameters were $\lambda =770$\,nm, $\Delta E_{kin}=\epsilon e^{-A}=75$\,meV, $\alpha=62°$, $\psi=0$, and $\sigma=1.815$\,ps. The ponderomotive momentum transfer is treated as momentum-dependent distortion to the kinetic energy axis. Yield-corrected reinterpolation generates the modulated ARPES signal. The results of this simulation depicted in Fig.~\ref{Sim_en_bunching} show that kinetic energy oscillations are becoming obvious at the Fermi edge because there the detected photoemission yield has a steep gradient.  At the two delays depicted in Fig~\ref{Sim_en_bunching} ($\tau= -4.25\,\mathrm{ps}$ and $\tau= -14.75\,\mathrm{ps}$) the oscillations at the Fermi surface are seen as transversal waves that propagate from the center outwards to higher absolute parallel momentum. The propagation velocity slows down with absolute parallel momentum which becomes visible in the decreasing wavelengths. Please note, that $\tau$ is defined in accordance with the convention of tr-ARPES by the difference in arrival times of the XUV and NIR pulses, that act as probe and pump pulses, respectively:
\begin{equation}\label{eq:def_tau}
\tau=t_{XUV} - t_{NIR}
\end{equation}

 A previously not described effect is the intensity modulations that are visible along curved bands. They are following lines of identical perpendicular momentum $k_{\perp}$ as given by Eq.~\ref{k_Pythagoras}. These modulations are visible for both presented delays. However, the modulations become stronger with later delays. In contrast to the modulations at the Fermi surface no gradient in the photoemission is needed to increase or decrease the electron yield in a certain energy interval. The intensity modulations become the dominant feature in parts of the ARPES spectrum that are showing no or only a strongly broadened band structure.\par \bigskip


\begin{figure}[tbhp]
	\centering
	\includegraphics[width=\linewidth]{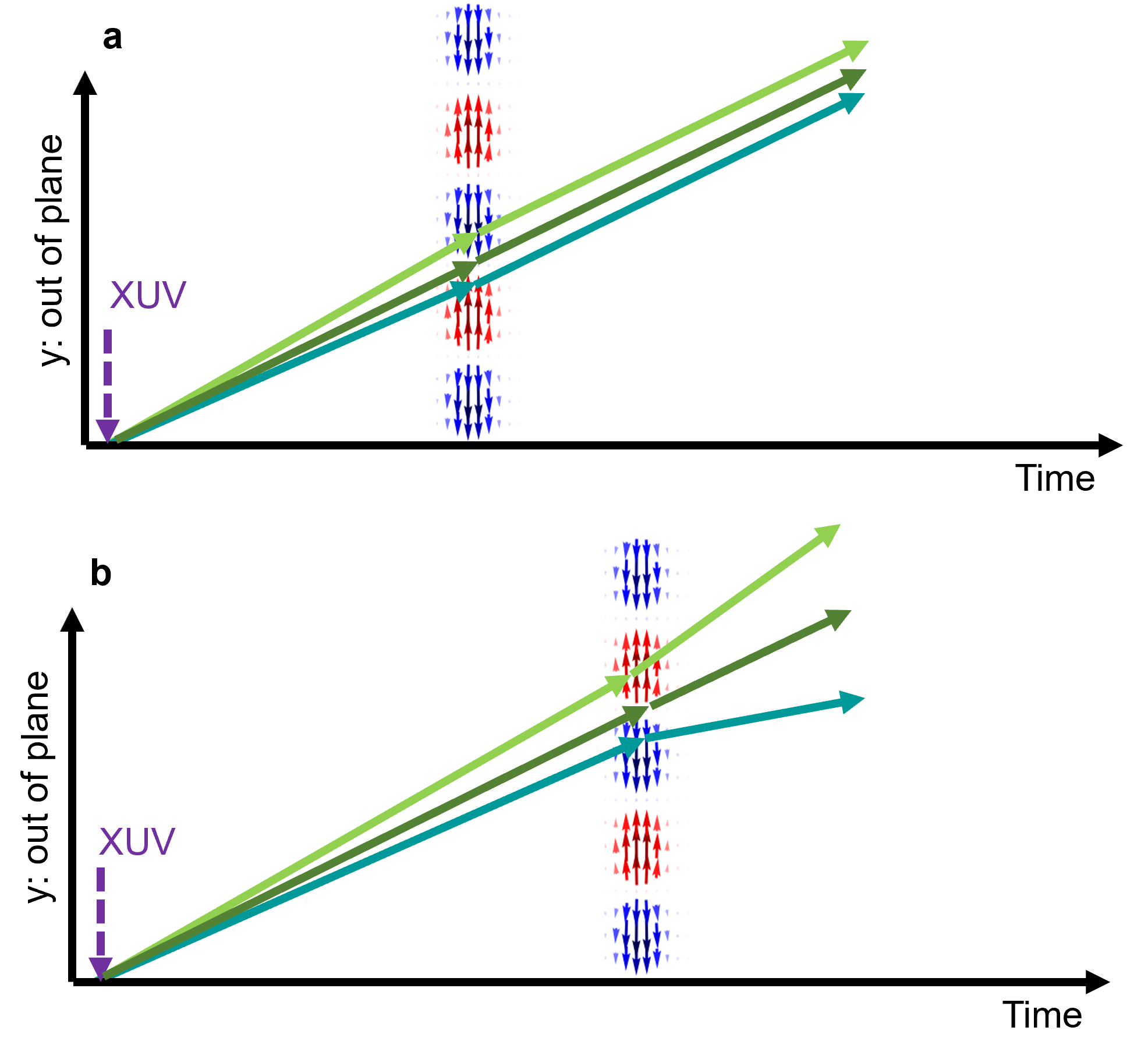}
	\caption{Sketch of electron energy bunching (\textbf{a}) and  anti-bunching (\textbf{b}). Electrons are generated at identical times but with different $v_\perp$ and are at different positions when laser field arrives. After momentum transfer $v_\perp$-distribution narrows (\textbf{a}) or broadens (\textbf{b}).}
\label{Explanation_en_bunching}
\end{figure}

Figure~\ref{Explanation_en_bunching} presents the basic mechanism behind the energy bunching. Electrons with identical parallel momentum but different perpendicular moment differ in kinetic energy and perpendicular velocity. As a consequence, even though they share the same starting position and time, these will disperse during propagation. Electrons with higher speed will travel further and will experience a different momentum transfer than slower ones. With increased optical delay between XUV and pump pulse this effect will initially increase. In the frequency picture electrons with different perpendicular velocity experience a difference in momentum transfer oscillation frequency $\Delta\Omega=\frac{2\pi}{\Lambda}*\Delta{v_\perp}$ which leads to a dephasing with delay. When the electrons still share the same valley in the ponderomotive potential, the momentum difference between the electrons will be decreased by the interaction with the laser pulse. The electron momentum distribution is reshaped by squeezing it onto a narrower momentum interval and resulting into increased electron yield around that momentum corresponding to the minimum of the valley. Conversely, at the peaks of the ponderomotive potential differences in electron velocities will be further enhanced effectively decreasing observed differential photoemission yield. At the spatial minima and maxima of the ponderomotive potential no momentum transfer occurs. Hence the effect of bunching and anti-bunching is strongest at delays corresponding to $n$ and $n+1/2$ full momentum transfer oscillation periods, respectively. The energy bunching effect gains in strength with increasing delay, as the electron distributions broadens during the propagation and electrons with a given difference in momentum will experience a increasingly divergent total momentum transfer. However, limits to energy bunching are ever smaller momentum intervals that are squeezed together as well as the laser pulse duration that will lead to an overall decrease of $U_p$ over delay.\par
Transforming the photoemission yield into a pure momentum space representation with axes $k_{\parallel}$ and $k_{\perp}$ the effect would take place independent from $k_{\parallel}$. Hence, bunching would ultimately structure the electron yield into flat discs perpendicular to the $k_{\perp}$ axis. \par
It should be noted that in ARPES only a small number of electrons per shot are leaving the sample. Otherwise Coulomb interaction will take over and electron spectra will be distorted. \textit{Bunching} is only experienced in the aggregated counts over a large number of single events after detection.\par 
Based on our simulations a review of results in \cite{bovensiepen_ultrafast_2009} is prudent. Observing intensity variations it was possible to map the dependence of the oscillation period on the electron kinetic energy. But the assumption that those could be explained with a remaining gradient of the PES does not hold true. In the data an increase of the oscillation amplitude with delay, which we now understand to be the onset of energy bunching, is visible.\par\bigskip
Instead of using a uniform photoemission density arbitrary ARPES data can be inserted into the routine. In Section \ref{Experimental_results} we are using measured ARPES spectra and show that experimental observations can be fully matched with simulations.\par

\section{\label{experiment}Experimental details}

A detailed description of the XUV beamline and other parts of our experimental setup can be found in Ref.\,\cite{frietsch_high-order_2013}. Our system operating at 10 kHz produces XUV photons in the range of 22 to 45\,eV with a pulse duration of approximately 50 fs through high harmonic generation. The fundamental NIR pulse with a photon energy of 1.59\,eV (780\,nm, 50\,fs) was employed for providing the transient grating. The 17th order high harmonic signal with a photon energy of 27.03 \,eV was selected using a toroidal grating monochromator. Photoelectrons were recorded with a hemispherical electron analyzer with acceptance angle of $\pm$\,\ang{15} in single electron counting mode (count rate of $\sim 0.5$ counts per laser pulse). XUV and fundamental beam were almost collinear and are overlapped spatially using the fluorescence of a thin Ce:YAG crystal on our ultrahigh vacuum (UHV) manipulator.\par By rotation of the sample holder the incidence angle of the optical pulses as well as the angle between sample normal and electron optics are manipulated simultaneously. This allows to observe ARPES signals at higher parallel momentum. Furthermore, it enables to vary the modulation wavelength of the transient standing light field (see Eq.~ \ref{Lambda_standing_Wave}).\par

The single-crystalline Fe(110) film of 3 nm thickness were grown in situ via molecular beam epitaxy on a W(110) substrate under UHV at room temperature and an evaporation pressure of $3 \cdot 10^{-10}$\,mbar. The 15 monolayer (ML) iron film exhibited a sharp low-energy electron diffraction pattern after annealing to 530\,K as expected for the Fe body-centered cubic (bcc) (110)-surface \cite{gradmann_periodic_1982} . Before starting the measurement the Fe film was magnetized in-plane by a magnetic field pulse of 100\,mT and cooled with liquid nitrogen to 100\,K. \par

\section{\label{Experimental_results} Experimental Results And Analysis}



\begin{figure}[tbhp]
	\centering
	\includegraphics[width=\linewidth]{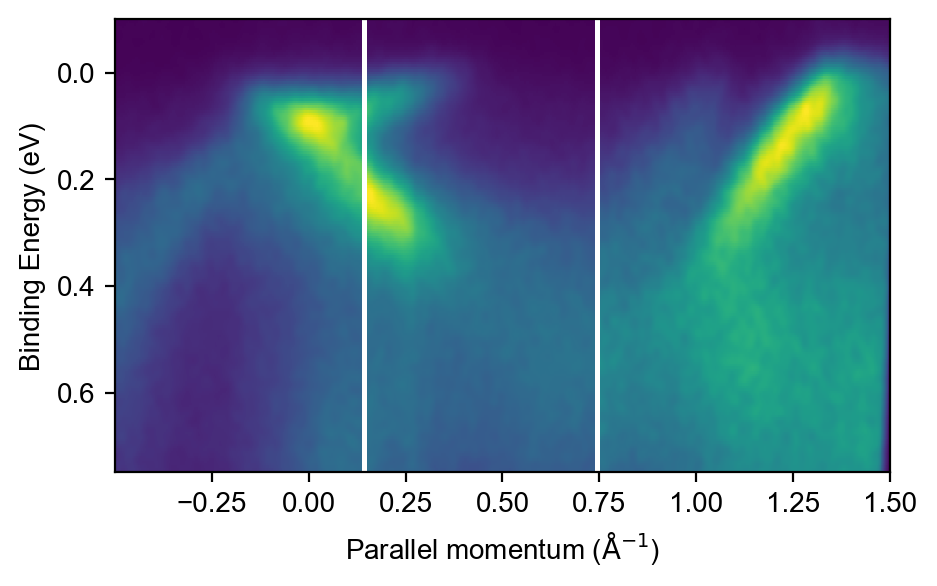}
	\caption{Composite ARPES spectrum of Fe. Increased range in parallel momentum was obtained by rotating the inclination of the sample. Spectra of each rotation angle was normalized individually. Electronic bands cut Fermi energy at $k_{\parallel}$ values of 0.3\,\AA, 1.1\,\AA, and   1.4\,\AA. 15° off H towards N.}
	\label{Mom_Fe_static_extended}
\end{figure}

\begin{figure*}[tbhp]
	\centering
	\includegraphics[width=\linewidth]{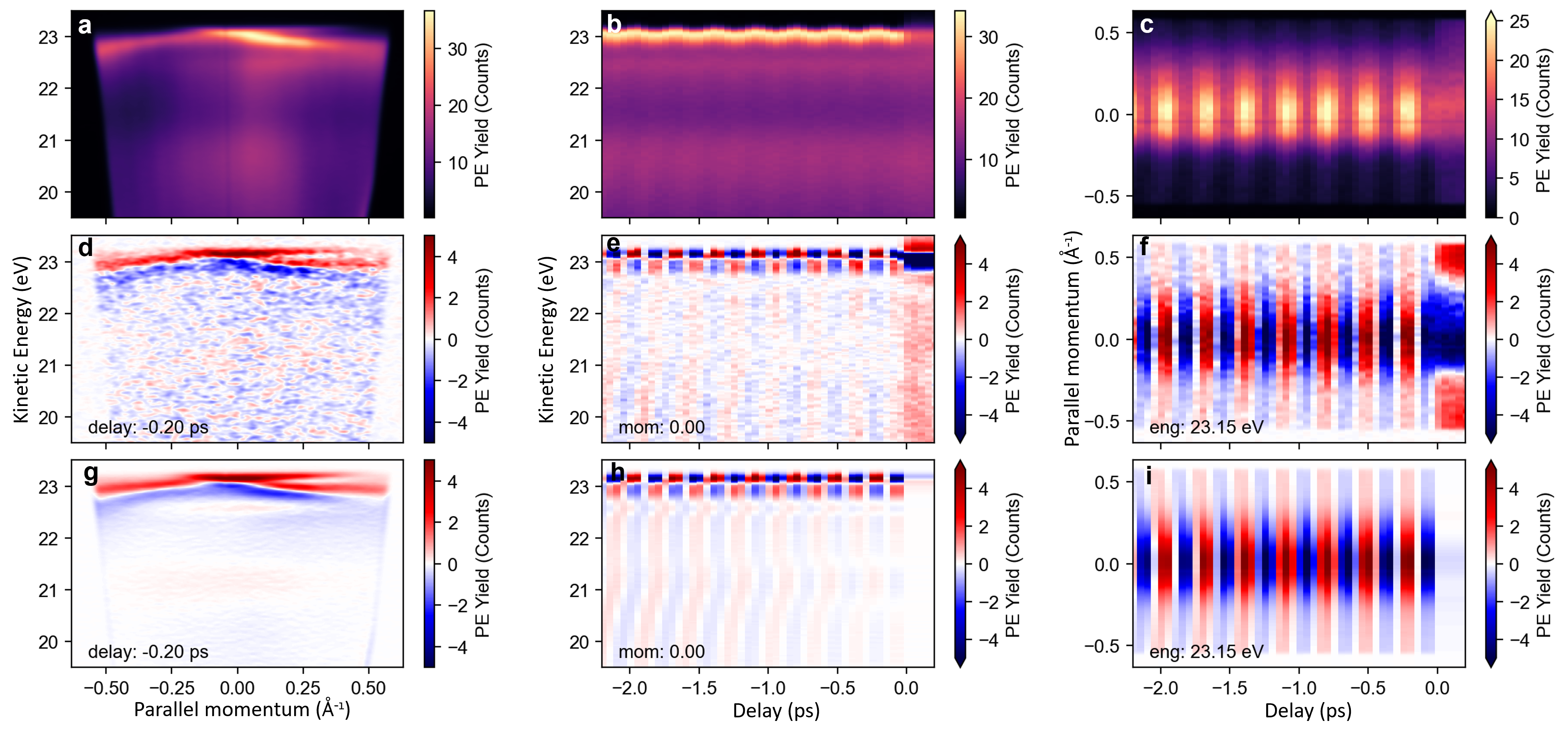}	
    \caption{Ponderomotive oscillations close to Brillouin zone center $\Gamma$ with sample surface aligned normal to the electron optic axis. Laser incidence angle $\alpha=$\,\ang{62}.  Averaged ARPES spectrum (\textbf{a}), Electron yield at fixed $k_\parallel$ (\textbf{b}) and fixed kinetic energy (\textbf{c}). (\textbf{d})-(\textbf{f}) Difference to mean electron yield with fixed delay, momentum, and energy, respectively. (\textbf{g})-(\textbf{i}) Fitted difference to mean electron yield with fixed delay, momentum, and energy, respectively. Best fit parameters were $\lambda =780.2$\,nm, $\Delta E_{kin}(E_{\mathrm{F}})=25$\,meV, $\psi=0.062$, and $\sigma=21.97$\,fs.}
	\label{Mom_Fe_Gamma}
\end{figure*}


We present angle-resolved photoemission spectroscopy on Fe(110) surface in the 2nd Brillouin zone. To extend the momentum range in our static measurements we rotate the inclination of the sample. The combination of three angle resolved photoemission spectra is depicted in Fig.~\ref{Mom_Fe_static_extended}.  Despite having a slightly tilted azimuthal angle of 15° with respect to the $\Gamma$-H direction, we obtain a cut through the Fermi surface which is in good accordance to the data presented in \cite{schafer_fermi_2005}. In Fe only bands close to the Fermi surface are visible due to strong electron correlation. At binding energies below 0.6\,eV the electronic band structure is broadened greatly. Electronic bands cut the Fermi level at $k_{\parallel}$ values of 0.3\,\AA, 1.1\,\AA, and  1.4\,\AA. Additionally, at $\Gamma$ overlap of electronic bands creates a high electron density at a binding energy of 0.09\,eV. At those features the strongest changes of the photoemission yield due to ponderomotive oscillations are expected.  \par


In our pump-probe experiment, the sample is pumped by the 780\,nm IR pulse at an incidence angle of $\alpha=$\,\ang{62} with an incident pump fluence of 11.7\,\unit{mJ/cm^2}. The photoelectrons are accelerated by the ponderomotive potential of the standing wave, which is generated by the IR pump pulse. Our datasets map the photoemission yield on a three-dimensional grid containing the kinetic energy, parallel momentum and delay axes. Figure~\ref{Mom_Fe_Gamma} (\textbf{a})-(\textbf{c}) displays the direct experimental yield, the yield difference of the obtained time-resolved data with respect to the time averaged spectrum in the measurement (Fig.~\ref{Mom_Fe_Gamma} (\textbf{d})-(\textbf{f})) and from fit results (Fig.~\ref{Mom_Fe_Gamma} (\textbf{g})-(\textbf{i})). The left column contains a delay averaged ARPES spectrum (\textbf{a}) as well as electron yield differences at delay $\tau=-0.2\,\mathrm{ps}$ in the measured data (\textbf{d}) and resulting from a fit (\textbf{g}).\par 
The ARPES spectrum presents a larger kinetic energy range than Figure~\ref{Mom_Fe_static_extended}. The observed electron yield at higher binding energy around $k_\parallel$ are in good accordance with \cite{Sanchez-Barriga2012}. The measured yield difference at $\tau=$-0.2\,ps (\textbf{b}) clearly shows that the states at the Fermi level have been lifted to higher energies generating an excess signal above the previous Fermi level and reducing the yield below. Smaller changes can still be observed at binding energies of 1-3 eV.\par
The center column (\textbf{b},\textbf{e},\textbf{h}) presents a cut through the data with the momentum fixed at the Fe-$\Gamma$ point for variable kinetic energies and delay. Here the oscillations are visible as delay dependent energy modulation of the Fermi surface. Additionally, most prominent at binding energies below 2\,eV, an oscillation of the electron yield building up for increasing delays. Plots with extended delay range, that further support this claim, can be found in the Supporting Material (see Figs.~SM1, SM2 and SM3). This is a clear signature of momentum bunching (cf. Fig.~\ref{Sim_en_bunching}). The difference signal shows continuous wave fronts that are slightly tilted indicating a dependence of the oscillation frequency on the electron kinetic energy. A strong dynamic that will not be discussed here is visible for $\tau>0$. This is the onset of the actual reaction of the sample to the excitation with the pump, which is of interest in a different context. \par 
The right column (\textbf{c},\textbf{f},\textbf{i}) shows the electron yield just above the Fermi edge. In that kinetic energy interval clear electron yield oscillations are observed (\textbf{c}) . Accordingly also the difference plots simply show the increase and decrease of the signal. At high delays a curvature in the oscillations becomes visible. This is connected with the higher oscillation frequency at low parallel momentum.\par The whole 3-dimensional data set that is partially shown in Fig.~\ref{Mom_Fe_Gamma} was used to obtain the best fit parameters were $\lambda =780.2$\,nm, $\Delta E_{kin}(E_{\mathrm{F}})=25$\,meV, $\psi=0.062$, and $\sigma=21.7$\,fs. From $\sigma$ the full-width at half maximum (FWHM) of the laser pulse at the sample surface can be estimated as 51.7\,fs. As the number of data points is very large ($>22\cdot10^6$) and only 4 independent variables are fitted - $\alpha$ and $t_0$ can be determined separately - the fit unambiguously finds the remaining parameters with high precision. A very good agreement is obvious when comparing the measured and simulated yield difference. The complete dynamics of the experiment are recovered. To a high extend differences can be attributed to noise. Furthermore, confidence in the measured data and fitting procedure allowed us to identify a minor flaw in the momentum calibration of our hemispherical electron analyzer. After implementing a corrected calibration generated by different means we obtained the presented data and benefited from an improved detection system.\par\smallskip

\begin{figure}[tbhp]
	\centering
	\includegraphics[width=\linewidth]{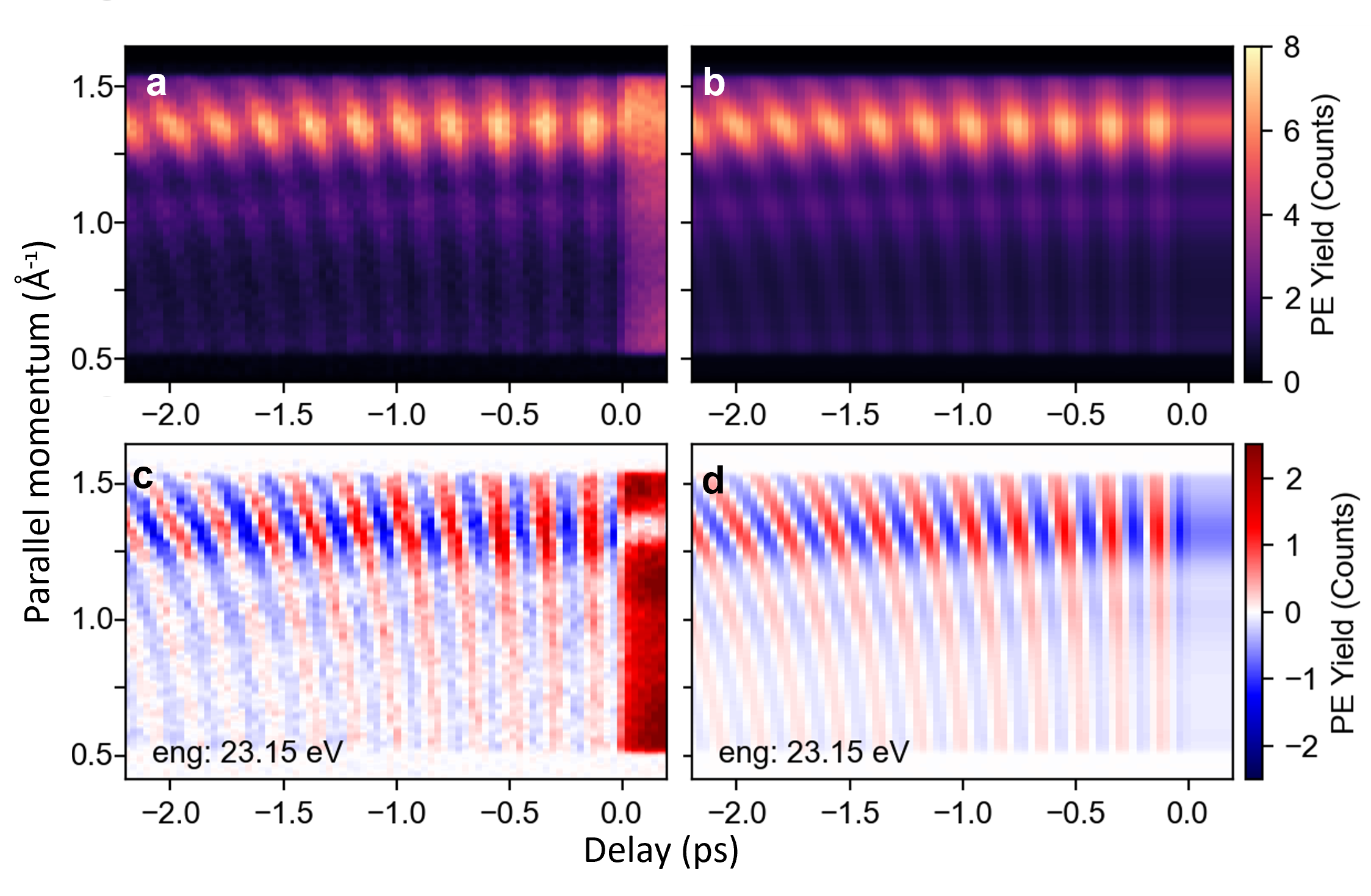}
	\caption{Ponderomotive oscillations observed at kinetic energies above the static Fermi level. Surface normal at angle of \ang{37.6} with respect to electron optic axis. Optical incidence angle \ang{24.4}. (\textbf{a}) Measured electron yield at fixed energy. The oscillation frequency drops with increased parallel momentum. (\textbf{b}) Difference to mean electron yield. (\textbf{c}) Fitted electron yield fully reconstructs the observed behavior before temporal overlap. (\textbf{d}) Difference to mean electron yield in fit. Best fit parameters were $\lambda =780.1$\,nm, $\Delta E_{kin}(E_{\mathrm{F}})=18$\,meV, $\psi=0.287$, and $\sigma=23.93$\,fs.}
	\label{Mom_Fe_high_mom}
\end{figure}

Fig.~\ref{Mom_Fe_high_mom} depicts data obtained with a rotation of the sample that reduced the incidence angle to $\alpha=$\,\ang{37.6}. Simultaneously this moved the axis of the electron optics of the hemispherical detector to \ang{24.4} from the surface normal. In the figure the kinetic energy is fixed to 23.15\,eV, which is above the static Fermi level. The reduction in $\alpha$ leads to an decrease of the light intensity modulation wavelength $\Lambda$ and, consequently, to an increase of the observed oscillation frequency $\Omega$ as described in eqs. (\ref{Lambda_standing_Wave}) and (\ref{Oscillation_freq}), respectively. Thus more oscillations are visible in the same delay interval as in Figure~\ref{Mom_Fe_Gamma}.  On the other hand, at higher parallel momenta the oscillations visibly slow down as the perpendicular momenta decrease as given by Eq.~\ref{k_Pythagoras}. Again the complete data set can be fitted without systematic deviation by adjusting geometry and pulse parameters.\par\smallskip

\begin{figure*}[tbhp]
	\centering
	\includegraphics[width=\linewidth]{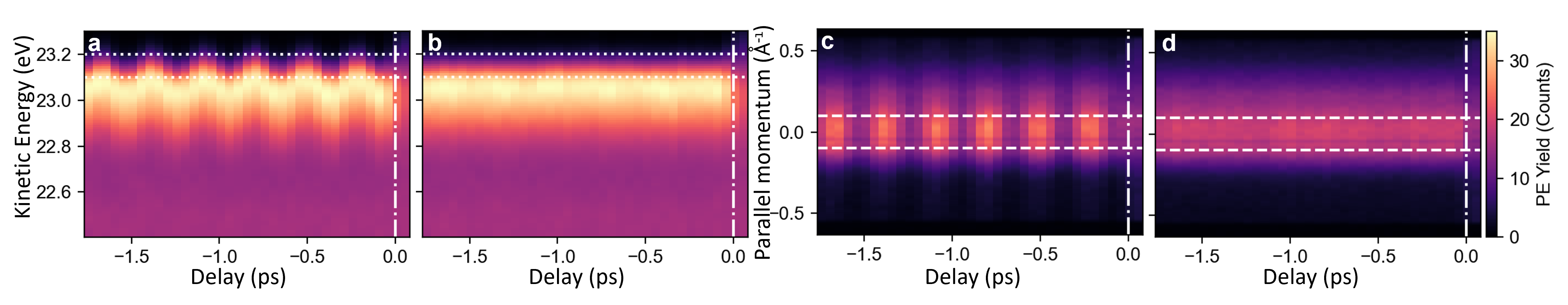}
	\caption{Reversal of effects of momentum transfer using fit parameters as input in simulation routine. Electron yield at fixed momentum as measured (\textbf{a}) and after energy correction (\textbf{b}). Electron yield at fixed energy as measured (\textbf{c}) and after energy correction (\textbf{d}). White lines indicate energy and momentum intervals used in the plots.} 
	\label{Mom_Fe_Gamma_reversed}
\end{figure*}

Fig.~\ref{Mom_Fe_Gamma_reversed} highlights the possibility to reverse the investigated momentum transfer in a tr-ARPES data set. After successfully fitting the parameter of the momentum transfer the numerical simulation routine can be employed to revert the influence of the pump pulse in each data point. This enables the analysis of tr-ARPES data performed with intense NIR pump conditions unperturbed by the effect of the ponderomotive field.\par

\section{\label{Conclusion} Conclusion}
As we have demonstrated the momentum transfer in the ponderomotive potential of ultrashort laser pulses affects tr-ARPES by modulation of the momentum perpendicular to the surface $k_\perp$. The strength and direction of the momentum transfer is determined by the properties of the transient standing light field as well as the propagation of the electrons before and during the interaction with the laser pulse. Consequences are energy shifts and also previously not described electron yield modulations in the ARPES signal. The energy bunching, that reshapes spectral features with increasing pump-probe delay is based on increased dephasing of the oscillatory momentum transfer in the continuous ARPES spectra. 
Critically, the route towards enhancing optically induced demagnetization in ARPES experiments is limited by space charge effects. To circumvent this the optical excitation has been pushed towards longer wavelengths. Yet the energy oscillations scale according to Eq.~\ref{Scaling_rel}, momentum transfer increases with the pump wavelength and fluence. Thus, the analysis of momentum transfer will have an increased importance in future ARPES measurements.\par
Increasing the photon energy ($\hbar\omega >100\,\mathrm{eV}$) to reach a higher Brillouin zone might be beneficial for certain ARPES investigations. Here the increased velocity of photoelectrons will lead to a shorter oscillation period, that may be shorter than the optical pulse widths. This would lead to a reduction of the visibility of the oscillations and rather to a broadening and reshaping of the ARPES spectra.\par

In this work we treated momentum transfer in the ponderomotive potential of ultrashort laser pulses on the basis of analytical solutions. Also the expansion towards quasi-continuous ARPES spectra utilized a parametrization corresponding to Eq.~\ref{BovenKineticOscillation}. A more general approach would include a full description of arbitrarily shaped ultrashort light fields and their reflection at selectable materials, the propagation of the resulting standing wave pattern along the surface and the calculation of the propagation of photoelectrons within the presence of a transient standing wave pattern. The solution would need to incorporate finite difference methods for calculating the electron trajectory as it is influenced by the transient field. This more detailed analysis allows for a prediction of the momentum transfer from basic principles. Dynamics at and after the temporal overlap of XUV and pump pulse can also be investigated with that approach. We are currently working on building a software repository and making it publicly available that will enable the treatment of ponderomotive momentum transfer under a broad range of experimental conditions.   

\begin{acknowledgments}
We acknowledge funding by the Deutsche Forschungsgemeinschaft (DFG, German Research Foundation) - Project-ID 328545488 CRC/TRR 227 \textit{Ultrafast Spin Dynamics}, project A01. We acknowledge fruitful discussions with Cornelius Gahl.
\end{acknowledgments}


\bibliography{references2}

\end{document}